\documentclass[11pt]{article}
\pdfoutput=1

\usepackage[utf8]{inputenc}
\usepackage{multirow}
\usepackage{amsmath, amsfonts, amssymb}
\usepackage{comment}
\usepackage{graphicx}
\usepackage{psfrag}
\usepackage{amsthm}
\usepackage[usenames,dvipsnames,svgnames,table]{xcolor}
\usepackage{enumerate}
\usepackage{arydshln}
\usepackage{soul}
 \usepackage{slashed}
\usepackage{subfig}
\usepackage{mathrsfs}
 \usepackage{a4wide}
  \usepackage{tikz}
  \usepackage{tikz-cd}
  \usetikzlibrary{shapes.geometric}
  \usepackage{tcolorbox}

  \usepackage{color}
  \definecolor{dark-gray}{gray}{0.20}
  \definecolor{gray}{gray}{0.30}
  \definecolor{light-gray}{gray}{0.80}
  \definecolor{dark-red}{rgb}{0.7,0,0}
  \definecolor{dark-green}{rgb}{0.1,0.4,0}
  \definecolor{dark-blue}{rgb}{0.3,0.3,0.7}
  \definecolor{light-blue}{rgb}{0.8,0.8,1}
      \definecolor{swamp}{RGB}{240, 199, 197}

  \usepackage{pifont}

\usepackage{newunicodechar} 
\newunicodechar{ỳ}{\`y}

\usepackage{setspace}

\newcommand{\be}{\begin{equation}}
\newcommand{\ee}{\end{equation}}
\newcommand{\eq}[1]{(\ref{#1})}

\def\be{\begin{equation}}
\def\ee{\end{equation}}
\def\bea{\begin{eqnarray}}
\def\eea{\end{eqnarray}}

\newcommand{\beq}{\begin{equation}}  \newcommand{\eeq}{\end{equation}}
\newcommand{\bal}{\begin{aligned}}   \newcommand{\eal}{\end{aligned}}
\def\beqa{\begin{eqnarray}}
\def\eeqa{\end{eqnarray}}

\def\simleq{\; \raise0.3ex\hbox{$<$\kern-0.75em
      \raise-1.1ex\hbox{$\sim$}}\; }
   \def\simgeq{\; \raise0.3ex\hbox{$>$\kern-0.75em
      \raise-1.1ex\hbox{$\sim$}}\; }

\numberwithin{equation}{section}

\usepackage{jheppub}
\usepackage{hyperref}
\usepackage{cleveref}

\hypersetup{
	colorlinks=true,
	linkcolor=dark-blue,
	citecolor=dark-red,
	urlcolor=dark-green,
	linktoc=page
}

\theoremstyle{remark}

\crefname{appendix}{Appendix}{Appendices}

\title{\centering The Desert and the Swampland}

\author{Cody Long$^1$,} \author{Miguel Montero$^1$,} 
\author{Cumrun Vafa$^1$,}
\author{Irene Valenzuela$^{1,2}$} \affiliation{Department of Physics, Harvard University, Cambridge, MA 02138, USA}
\affiliation{Instituto de F\'{i}sica Te\'{o}rica UAM-CSIC and Departamento de F\'{i}sica Te\'{o}rica, Universidad Aut\'{o}noma de Madrid, Cantoblanco, 28049 Madrid, Spain}

\emailAdd{c.e.cody.long@gmail.com}
\emailAdd{mmontero@g.harvard.edu}
\emailAdd{vafa@g.harvard.edu}
\emailAdd{irene.valenzuela@uam.es}

\abstract{The most natural expectation away from asymptotic limits in moduli space of supergravity theories is the \emph{desert scenario}, where there are few states between massless fields and the quantum gravity cutoff.  In this paper we initiate a systematic study of these regions deep in the moduli space, and use it to place a bound on the number of massless modes by relating it to the black hole species problem.  There exists a consistent sub-Planckian UV cutoff (the species scale) which resolves the black hole species problem without bounding the number of light modes. We reevaluate this in the context of supersymmetric string vacua in the desert region and show that even though heuristically the species scale is compatible with expectations, the BPS states of the actual string vacua lead to a stronger dependence of the cutoff scale on the number of massless modes.  We propose that this discrepancy, which can be captured by the ``BPS desert conjecture'', resurrects the idea of a uniform bound on the number of light modes as a way to avoid the black hole species problem.   This conjecture also implies a stronger form of the Tadpole Conjecture, which leads to an obstruction in stabilizing all moduli semi-classically for large number of moduli in flux compactifications.}

\begin{document}
\hypersetup{pageanchor=false}
\makeatletter
\let\old@fpheader\@fpheader
\preprint{IFT-UAM/CSIC-21-156}

\makeatother

\maketitle

\hypersetup{
    pdftitle={A cool title},
    pdfauthor={Cody Long, Miguel Montero, Cumrun Vafa, Irene Valenzuela},
    pdfsubject={Finiteness of the string Landscape, Calabi-Yau}
}

\emailAdd{mmontero@g.harvard.edu}
\emailAdd{irene.valenzuela@uam.es}
\emailAdd{vafa@g.harvard.edu}
\emailAdd{mmontero@g.harvard.edu}

\newcommand{\remove}[1]{\textcolor{red}{\sout{#1}}}

\newpage

\section{Introduction}
\label{sec:intro}

One of the principles of the Swampland program is that there are a finite number of quantum gravity backgrounds \cite{Vafa:2005ui,Acharya:2006zw,Douglas:2010ic,Hamada:2021yxy}.   More precisely if we fix a cutoff we expect a finite number of string vacua.
This in particular suggests that there is an upper bound on $N$ the number of massless or light species  modes allowed in a consistent quantum gravity theory.  A heuristic explanation of this may have been that if there were no upper bound on $N$ then the entropy of a fixed size black hole which should be at least $N$ (as the black hole can be made up of such constituents) and this would lead to a contradiction if there were no bound on $N$.
However, this reasoning is not quite correct.

In a quantum theory of gravity we expect that any effective theory breaks down for energies $E\geq M_{\text{P}}$, where quantum gravitational effects become strong and invalidate the low energy effective theory.
Therefore in any effective theory including quantum gravity we expect to have a cutoff $\Lambda \leq M_{\text{P}}$.  In some cases we expect $\Lambda \sim M_{\text{P}}$.  This is the case for example for M-theory in 11 dimensions.
On the other hand there are also cases where $\Lambda \ll M_{\text{P}}$.  For example for 10d string theory the cutoff is given by the string scale $\Lambda \sim M_{s}\sim g_s^{1/4} M_{\text{P}}$ which for $g_s\ll1$ is far below the Planck scale.

For a given cutoff, the radius of black holes which can be described reliably by the effective theory are bounded by the cutoff scale: $R_{BH}>1/\Lambda$.  In d dimensions, the entropy of such a black hole is thus bounded by
$S> (M_{\text{P}}/ \Lambda)^{d-2}$.  Since the entropy of such a black hole should be bigger than the number of light species $N$, we learn that if 
$N< (M_{\text{P}}/ \Lambda)^{d-2}$ there would be no contradiction with the fact that the black hole entropy should be more than the number of light species.  In other words if $\Lambda \leq \Lambda_{\rm sp}$ with the species scale $\Lambda_{\rm sp}=M_{\text{P}}\cdot N^{-\frac{1}{d-2}}$, there would be no contradiction between having an arbitrarily large number of light species $N$ and the finiteness of the black hole entropy \cite{Dvali:2007wp,Dvali:2007hz,Dvali:2010vm}.

The main goal of this work is to evaluate this resolution of the black hole species problem from the perspective of string landscape.  More specifically we focus on supersymmetric string landscape for which we have more analytic control. To test this proposal we need to investigate points on the landscape for which the number of light/massless modes can be large.  This happens when we have 8 or less supercharges in the theory.  The first such case is ${\cal N}=(1,0)$ supersymmetric theories in 6d.  We focus on the number of massless tensor multiplets $N$ and attempt to compute the cutoff $\Lambda(N)$.  First of all we need to be more precise what we mean by $\Lambda(N)$.  As we change the moduli of the supersymmetric theory by changing the vev of scalar fields, we expect that mass of excited states change, and so in particular the first massive state which is not part of the EFT which defines $\Lambda$ will depend on the massless moduli.  The first question is whether we can have $\Lambda$ become equal to the $M_{\text{P}}$?  In other words, can there be a {\it desert scenario} where there is no light states all the way till the Planck scale? This is interesting since the largest possible value of $\Lambda$ provides the strongest constraint on the number of light species $N$ via the species bound above.  To address this issue, we show that no matter what  moduli we choose, there is a $\Lambda < M_{\text{P}}$ for which there is a particle of that mass $m\sim \Lambda$ which is not part of the EFT. We do this by focusing on BPS states for which we can compute the mass exactly as a function of the moduli using supersymmetry.
  In particular we cannot quite have the desert all the way to the Planck scale.  The next question would be whether the $\Lambda$ we find scales with the number of massless modes as $N^{-1/(d-2)}$ as one would expect based on species scale.  Quite surprisingly we find, using supersymmetry and some genericity assumptions, that for theories in $d=4,5,6$ with 8 supercharges this behavior would naively be expected to hold.  However, upon more careful scrutiny by focusing on actual examples (a mini-Landscape of 34864  6d F-theory models with toric base) and focusing on BPS states we show that this genericity assumption seems false and that $\Lambda\sim N^{-a}$ with $a>1/(d-2)$.  At first site this may sound contradictory with the species scale.  However the $\Lambda$ we have computed is based on actual supersymmetric states.  However, there are potentially other lighter states in the theory which are not supersymmetric which can resurrect the dependence of species scale with $N$.  This explanation can work as long as $N<N_0$ for some upper bound $N_0$.  If $N> N_0$ since the slope of the supersymmetric cutoff $\Lambda$ with $N$ is bigger than  the species scale, the supersymmetric cutoff would not agree with the species scale.  We propose that this may explain why there is an actual upper bound in the number of massless modes for theories with 8 supercharges.  

The organization of this paper is as follows:  In Section \ref{sec:species-2} we discuss the general setup for the supersymmetric landscape.  In Section \ref{sec:16and32} we discuss the situation for theories with 16 or 32 supercharges.  In Section \ref{sec:heuristic} we give a heuristic argument why in theories with 8 supercharges the supersymmetric cutoff $\Lambda$ for theories with dimensions $d=4,5,6$ agrees with the species bound.  In Section \ref{sec:8}, however, we show this heuristic expectation based on genericity is incorrect by actually looking at a large class of examples for ${\cal N}=(1,0)$ supersymmetric theories in $d=6$ and offer our resolution of this.  We also discuss applications to the Tadpole Conjecture in this Section.  In Section \ref{conclus} we present our conclusions.

\section{The General Setup}\label{sec:species-2}
Let us consider a supersymmetric theory coupled to gravity.  Such theories typically have massless moduli fields, if there is enough supersymmetry.  In such a case a natural question to ask is whether the low energy description involving only the massless modes is adequate all the way to Planck energy.  In other words, is the first excited state of the order of Planck mass or higher?  If there is a large number of massless modes in the supersymmetric theory, we would expect this not to be possible because otherwise the Planck mass black holes (or slightly above it) would be reliably described by such an effective theory and would yield an entropy too small to account for all the massless modes that the black hole can be made of, violating the species bound explained in the Introduction. 

We know that the mass of excited states will depend on massless moduli $\Lambda(\phi)$.  
Note that we do not expect $\Lambda(\phi)$ to have a non-vanishing minimum, as there are always infinite distance points on the moduli where we expect exponentially light states in accord with the Distance Conjecture  \cite{Ooguri:2006in}.  On the other hand it is natural to ask how close to the Planck scale can we make the cutoff $\Lambda$ where the first excited particle appears by varying moduli.  In other words we are interested in finding a point on the moduli space of massless fields where the cutoff is maximized and we call this the $\Lambda$ of the supergravity theory.
$$\Lambda ={\rm max}_\phi \Lambda(\phi)$$
The difficulty with finding $\Lambda$ is that we do not have a reliable method to compute the excited spectrum of the supergravity theory.  However, there is a subset of states in supersymmetric theories that we control, namely the BPS states.  Such states have masses or tensions which are determined exactly by supersymmetry.  We can use this to find an upper bound on the value of $\Lambda$.  Namely if we define
$$\Lambda_{BPS}(\phi)={\rm min}_{BPS} \{ m_{BPS}(\phi)\}$$
by scanning over the BPS states, then we can define the desert among all the BPS states by considering
$$\Lambda_{BPS}={\rm max}_\phi \{ \Lambda_{BPS}(\phi)\}={\rm max}_\phi\{ {\rm min}_{BPS} \{ m_{BPS}(\phi)\}\}$$
where $\Lambda_{BPS}(\phi)$ denotes the first massive BPS state a a given point $\phi$ in moduli.  It is clear from the definition that the actual $\Lambda$ is bounded by $\Lambda_{BPS}$:
$$\Lambda \leq \Lambda_{BPS}.$$

Here we have to be a bit more careful about what we mean by the mass associated to a BPS object.  For a BPS particle it is simply the mass of the state.  However, for BPS p-branes whose tensions $T^{BPS}_p$ is determined using the BPS algebra, we need to form localized particle states which would necessarily be non-BPS.  Such particle states which are themselves non-BPS states but come from BPS objects will have mass scales of the order of $m_{BPS}\sim (T^{BPS}_p)^{1/(p+1)}$.  This is of course rather familiar in the context of fundamental strings.  Therefore in the above for each BPS state what we have in mind is actually
$$m_{BPS}\sim (T^{BPS}_p)^{1/(p+1)}$$
In other words we have
$$\Lambda_{BPS}={\rm max}_\phi\{ {\rm min}_{BPS_p}\{ (T^{BPS}_p)^{1/(p+1)}(\phi)\} \}$$

Throughout this paper, we will compute $\Lambda_{BPS}$ for different theories, organizing the discussion according to the level of supersymmetry. For theories with 8 supercharges, the Landscape is rich enough to study the dependence of $\Lambda_{BPS}$ with the number of massless modes, i.e. $\Lambda_{BPS}(N_{\rm massless})$. In that case, we can also compare the results with the species scale \cite{Arkani-Hamed:2005zuc,Dvali:2007wp,Dvali:2007hz,Dvali:2010vm},
\beq
\label{species}
\Lambda_{\rm sp}\sim M_{\text{P}}\cdot N^{-{\frac{1}{d-2}}}
\eeq
where $N$ is the number of weakly-coupled species with a mass at or below $\Lambda_{\rm sp}$. Notice that, in this definition, $N$ are not only the massless modes but include also all the light modes that can be described within the EFT below this quantum gravity cut-off. In fact, near the infinite distance limits of the moduli space, this species scale is dominated by the infinite towers of states becoming light \cite{Ooguri:2006in,Grimm:2018ohb,Heidenreich:2017sim}. However, in the deep interior of the moduli space where the BPS mass gap gets maximized, we have a quasi-dessert scenario in which the asymptotic towers of states are very heavy and the species bound will instead be dominated by the massless modes. In other words, in the interior of the moduli space, we expect to have, in general, many more massless modes than light massive modes below $\Lambda_{\rm sp}$. In that case, it should be a good approximation to replace $N$ by $N_{\rm massless}$ in \eqref{species}, as we will do when comparing our results with the species bound in Section \ref{sec:8}.

\section{BPS mass gap for theories with more than 8 supercharges}\label{sec:16and32}

We will now analyze the maximal gap predicted by the BPS sector in theories with 32 and 16 supercharges, where complete calculations are possible. Notice that we will use BPS masses and tensions as a proxy of the actual mass gap; therefore, what we obtain is actually an upper bound on how large the mass gap can be made in these higher-dimensional moduli spaces. Due to the simplicity of the setups involved, we also have some degree of control for some non-BPS states. 
\subsection{32 supercharges}
The first and easiest example to analyze is the 11-dimensional vacuum of M-theory. The massless fields include no moduli; there is no optimization to perform, and the mass gap set by BPS states is the Planck scale. More precisely, if we estimate the gap from the BPS sector, there are mass scales associated to the M2 and M5 branes, which are
\begin{equation} M_{\text{M2}}\sim T^{1/3}_{\text{M2}},\quad M_{\text{M5}}\sim T^{1/6}_{\text{M5}}.\label{ext3}\end{equation}
These two quantities are of order the 11d Planck mass. There is an intrinsic $\mathcal{O}(1)$ ambiguity in dxetermining a mass scale out of the tension of an extended object, which explains the twiddles in \eq{ext3}. We will face similar issues in most of the lower-dimensional vacua we will analyze in this paper.

Compactifying M-theory on a circle we reach IIA string theory. There is now a modulus, the dilaton, which controls the ratio between string and Planck scales. The tensions of D-branes, NS5-brane, and the fundamental string itself, are controlled by different powers of $g_s$. Recall the formula for  the tension of a Dp-brane in Planck units \cite{polchinski1998string},
\begin{equation}T_{\text{Dp}}=g_s^{\frac{p-3}{2}}.\label{tdp9}\end{equation}
 The point of largest gap is then reached when $g_s\sim 1$, and all scales coincide with the Planck scale. Not surprisingly, we find that the point of largest gap is as far away from the infinite distance limits of moduli space as one can be; as we will see in later Sections, it is also as far as it can be from any finite-distance singularity where charged states become light. From \eq{tdp9}, we also note that since the tensions of BPS objects are determined by their couplings, pairs of dual D-branes satisfy
 \begin{equation}T_{\text{Dp}}\cdot{T_{\text{D(6-p)}}}\sim \text{independent of $g_s$}.\end{equation}
What this means is that, as one object gets heavy, the electromagnetic dual becomes light, so the natural point where both electric and magnetic objects are heavy is in the middle of moduli space where $g_s\sim 1$.  This corresponds to compactifying M-theory on a circle with radius of Planck length.  Notice that at the extremum there are more than one object with the same mass scale.  This pattern will be repeated in all the examples below.

Lower-dimensional toroidal compactifications of M-theory follow the same pattern, with the BPS spectrum obtained from gravitational solitons such as KK monopoles, and M2 or M5 branes wrapped on the same cycles. The point of maximal gap is then a Planckian torus, with all sides similar to each other. 

The one remaining higher-dimensional vacuum is 10-dimensional IIB. There is a complex modulus, the axiodilaton $\tau$, and the spectrum of extended objects includes $(p,q)$-strings and five branes, $D3$-branes, as well as $(p,q)$ 7-branes. The tension of the $D3$ brane in Planck units is $\mathcal{O}(1)$ and independent of $\tau$, so it does not play a role in our discussion. We will also ignore $(p,q)$ 7-branes, since, due to their long-range interactions, the mass scale obtained from their tension is not a good proxy for the mass gap of the theory. This leaves $(p,q)$ strings and 5-branes as the objects setting the gap. The formula for both of them is very similar, taking the form \cite{polchinski1998string}
\begin{equation} T_{(p,q)\text{ string}}= \frac{\vert p + q\tau\vert}{\sqrt{\tau_2}}\label{eq252}\end{equation}
in Planck units for strings, and a similar expression but with $\tau$ replaced by $-1/\tau$ for the $(p,q)$ 5-branes. Similarly to the discussion around \eq{ext3}, assuming the gap is at the Planck scale or below, it will be set by the $(p,q)$-strings and not the 5-branes, since in obtaining a mass scale one takes a fifth root rather than a square root for the latter versus the former. We are led to the problem of maximizing the minimum of \eq{eq252} over $(p,q)\in \mathbb{Z}^2$ as a function of $\tau$. This problem has a simple geometric interpretation: $\tau_2$ is the volume of the lattice in $\mathcal{C}$ generated by $\{1,\tau\}$, and so $\vert p + q\tau\vert$ is the length of a lattice vector. By rescaling the lattice vectors by a real number $\lambda$, we can make sure that the shortest lattice vector has norm 1. The problem then becomes finding the maximal value of the volume $\lambda^2\tau_2$ of the fundamental cell of the lattice, subject to the constraint that the minimal length vector has volume one. By drawing a sphere of diameter 1 around each lattice point in such a lattice, one obtains a sphere packing. Thus, the problem we have is equivalent to the classic problem of sphere packing \cite{2010arXiv1009.4322C}, which in two dimensions is solved by an hexagonal lattice. We conclude that the optimal point, with the maximal gap for $(p,q)$-strings, corresponds to the hexagonal lattice with $\tau$ a third root of unity. Although this corresponds to one of the special ``cusps'' in the torus complex structure moduli space, it is interesting that it does not correspond to the cusp at $\tau=i$, as one might have naively expected. In any case, the gap we obtain is again $\mathcal{O}(1)$ in Planck units, and corresponds to any of the fundamental string, the D1 brane, or the $(1,1)$ string, all of which have the same tension.

\subsection{16 supercharges}\label{sec:16}
We now analyze theories with 16 supercharges. This class of theories includes minimally supersymmetric models in seven, eight, and nine dimensions. Unlike theories with 32 supercharges, for which only one massless multiplet (the gravity multiplet) is possible, theories with 16 supercharges can have both gravity and vector multiplets \cite{Aharony:2007du}. Low-energy interactions are fully determined by supersymmetry, to the point that the one free parameter in the theory is the number of vector multiplets at a generic point in moduli space, known as the rank of the theory \cite{Aharony:2007du,Ortin:2015hya}. Theories with 16 supercharges have been the subject of much recent work from the Swampland perspective \cite{Kim:2019vuc,Montero:2020icj,Dierigl:2020lai,Cvetic:2020kuw,Hamada:2021bbz,Bedroya:2021fbu}, to the point that Swampland principles are enough to ``bootstrap'' the known string theory compactifications, realizing the String Lamppost Principle. In particular, there is an upper bound on the rank of the theory, which is $26-d$ for a $d$-dimensional theory \cite{Kim:2019ths}.

We will now partly address the question of the maximal gap in the BPS spectrum for toroidal compactifications of heterotic string theory, which will illustrate the general features of the problem we will study later. The reason we will not do an in-depth study of these compactifications is that the Landscape of these models is finite, and not very large; for instance, in nine dimensions, the moduli space has four connected components only \cite{Aharony:2007du}. This means that we cannot increase the number of massless fields arbitrarily, and therefore, we do not have enough statistics to meaningfully study the dependence of the gap with the number of fields. This problem will be remedied in Section \ref{sec:8}, by reducing the number of supercharges.

Since the theory contains vector multiplets, a natural set of charged objects to consider are particles, which can in fact be BPS \cite{Aharony:2007du}. The advantage of a perturbative heterotic setup is that we have control over all perturbative states, both BPS and non-BPS, so that we can also investigate (in perturbative corners) to what extent the BPS states give a good estimate of the actual mass gap. Of course, deep in moduli space, only the masses of BPS states are reliable. 
 
The well-known perturbative mass formula in string units is \cite{Font:2020rsk}
\beq
m^2=\frac{1}{\alpha'}\left(p_L^2+p_R^2+2(N+\bar N-1)\right)
\eeq
where $p_{L,R}$ are the left-(right-)handed worldsheet momenta and $N(\bar N)$ the left(right)-handed oscillation number. Upon imposing the level-matching condition
\beq
p_L^2-p_R^2=2-2(N-\bar N),
\label{level}
\eeq
the mass spectra reads
\beq
\label{mass16}
\alpha'\, m^2=2p_R^2+4\bar N=2p_L^2+4N-4\ .
\eeq 
BPS states, preserving half the supercharges, form a cone, and in heterotic string theory, correspond to states having $\bar N=0$ \cite{PhysRevLett.63.478}. This means that
\beq
\alpha'\,  m_{BPS}^2=2p_R^2,
\eeq
and must satisfy $p_L^2-p_R^2\leq 2$ due to \eqref{level}. We can distinguish three types of BPS states:
\begin{itemize}
\item $p_L^2-p_R^2= 2$: Field theory states. They can becomes massless at finite distance and induce an enhancement of the gauge group.
\item $p_L^2-p_R^2= 0$: These towers of states can only become massless at infinite distance.
\item $p_L^2-p_R^2< 0$: Black hole states. Their mass in string units remains bounded below by an $\mathcal{O}(1)$ number everywhere in moduli space, as follows from \eq{mass16} and the fact that $N> 1$ for these states (from \eq{level}).
\end{itemize}

Intuitively, we expect that the maximum gap will occur at a regular point in the interior of the moduli space, and as far away as possible from all singularities including the gauge field theory enhancements. In this sense, it is the \emph{deepest} we can be in the moduli space. Only states with $p_L^2-p_R^2\geq 0$ can become lighter than the string scale, so they will be the relevant states in setting the gap $\Lambda_{BPS}$. Contrarily, the black hole states will always be more massive and play no role in determining this mass gap. 

To give an illustrative example, we will now discuss the case of a circle compactification of heterotic string theory (nine dimensions) without Wilson lines. 
We can view this as keeping $E_8\times E_8$ as part of the massless modes, leaving only the radius (in addition to string coupling constant) as the moduli to vary.  The momenta are then given by \cite{Font:2020rsk}
\beq
p_{L,R}=\frac1{\sqrt{2}R} \left(n \pm R^2 w\right)
\eeq
where $n,w$ are the KK and winding number, while $R$ is the circle radius in string units. Moreover, BPS states satisfy $p_L^2-p_R^2=2nw\leq 2$. Indeed, the asymptotic towers of states (either purely KK or purely winding modes) have $p_L^2-p_R^2=0$, while the only field theory state here has $n=w=1$ and becomes massless at the self-dual point $R=1$. 

The mass formulae above are given in string units. For Swampland purposes, it is more natural to write masses in nine-dimensional Planck units, which are related to the above via a rescaling by a factor depending on the 9d dilaton.
This implies that the masses also depend on $\phi$, the heterotic dilaton, in a way that is independent of the charges. The BPS spectrum masses can grow without bound; at strong coupling, where the system is described by type I' string theory \cite{Bergman:1997py}, BPS states correspond to D0-D8 bound states, and can become arbitrarily massive, being far above the actual gap of the theory, which is set by non-BPS fundamental type I' string states. For this reason, the spectrum of BPS states is only a good proxy for the actual gap of the theory at heterotic coupling constant of order 1 and below. Fortunately, it is precisely at $g_s\sim \mathcal{O}(1)$ where we expect to find the maximum gap of the theory, by a similar discussion as in the previous Section.

\begin{figure}[!htb]
\centering
\includegraphics[width = 0.5 \textwidth]{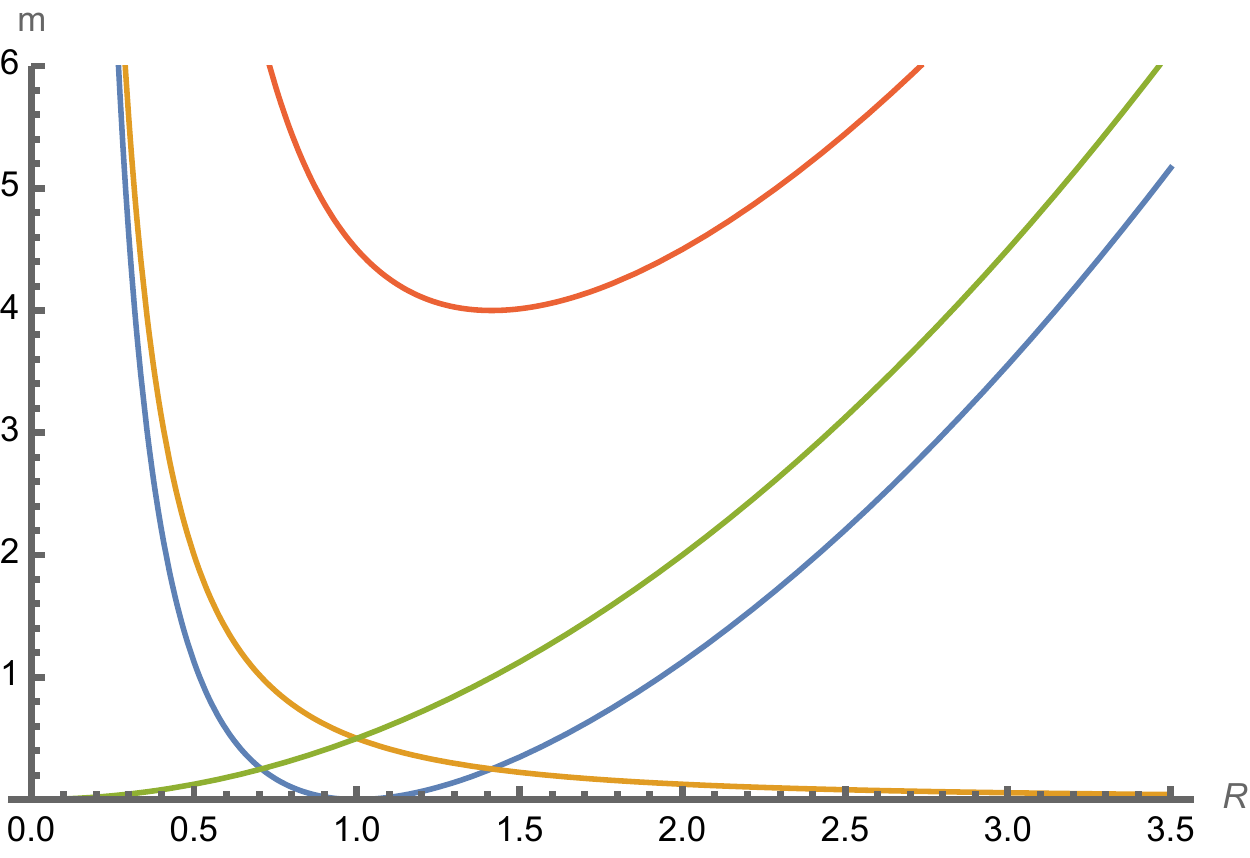}
\vspace{0.3cm}
\caption{Mass behavior (in string units) of different BPS states in terms of the circle radius for a 9d theory with 16 supercharges (heterotic on $S^1$ with Wilson lines switched off).
The blue, red, yellow and green curves denote masses of field theory, black hole, and the two infinite distance light towers respectively.}
\label{fig16}
\end{figure}

The mass behavior of the different states is plotted in Figure \ref{fig16}. The maximum mass gap occurs at the point at which the mass for field theory state crosses the mass of the towers. As one can read from the plot, this occurs at $R=\frac1{\sqrt{2}}$ and yields $\Lambda_{BPS}=0.25M_s$.

It is also interesting to check the mass gap that arises when including the field theory states of $SU(2)$ as part of the EFT. This will then freeze the radius at $R=1$, and the only moduli left is the string coupling constant which we expect to lead to $g_s\sim 1$ for maximal gap of KK (or dual winding) towers.

Due to \eq{mass16}, non-BPS states ($\bar N\neq0$) have a mass lower-bounded by 4 in string units. Hence, the gap will be set by BPS states, at least perturbatively, as long as there is always a BPS state with mass less than 4, or equivalently, with $p_R^2\leq 2$. This is automatically true in the restricted sections of the moduli space of toroidal compactifications of the heterotic string that we just analyzed.

We will see that, also in theories with 8 supercharges, $\Lambda_{BPS}$ will be determined by field theory states, and the maximum gap will occur at the point which is furthest away from all finite and infinite distance singularities. The asymptotic towers play the leading role only when field theory states are not included.

As emphasized above, we did not carry out a systematic analysis of theories with 16 supercharges because the corresponding Landscape is too small to provide us with any interesting dependence on the number of moduli. We will nevertheless finish with some general comments on how the analysis would be carried out. Just like in heterotic, the natural BPS objects to consider are particles. For a rank $r$ theory, these charged states live naturally in a charge lattice $\Lambda_{10-d,10-d+r}$, which is equipped with an inner product of signature $(10-d,10-d+r)$. All these theories can be described perturbatively in corners of their moduli space, and in these corners BPS states are naturally Dabholkar-Harvey states \cite{Font:2020rsk} just as in heterotic. Then the analysis above can be repeated.

\section{Species bound and mass gap in Calabi-Yau compactifications}\label{sec:heuristic}

Let us next reduce the level of supersymmetry and consider compactifications with eight supercharges. This includes 6d ${\cal N}=$1, 5d ${\cal N}=$1 and 4d ${\cal N}=$2 theories arising from Calabi-Yau (CY) threefold compactifications. From now on, we are going to focus only on the maximum  mass gap generated by BPS states, since the full non-BPS spectrum is not well enough known to allow for a systematic analysis.

In general, the mass scale of a (p+1)-dimensional BPS state coming from a  brane wrapping a cycle $\Sigma$ is given by
\beq
\label{BPSmass}
m_{BPS}=M_D(vol_\Sigma)^{1/(p+1)}=\frac{M_d}{\mathcal{V}^{1/(d-2)}}(vol_\Sigma)^{1/(p+1)}
\eeq
where $vol_\Sigma$ is the volume of the wrapped cycle and $M_D$ is the higher dimensional Planck mass for M-theory compactifications or the string scale in the case of string compactifications. Furthermore, $M_d$ is the lower dimensional Planck mass and the overall volume $\mathcal{V}$ is written in $M_D$ units. Depending on the space-time dimension, the lightest BPS object will be a particle or a string. 

The goal is to find CY moduli where for all BPS states we can push up their mass to the maximum value. Hence, we want to make the volume of all cycles $\Sigma$ as large as possible while keeping the overall volume $\mathcal{V}$ fixed to one.  In principle one may think we can get this as large as we wish by sending $\mathcal{V}\rightarrow 0$, however, it turns out that $(vol_\Sigma)^{1/(p+1)}$ scales the same way as $\mathcal{V}^{1/(d-2)}$ for all the cases of interest here, so we can set $\mathcal{V}=1$ with no loss of generality.  The independence of the mass in Planck units from the volume follows from the decoupling of hypermultiplets (which includes the volume moduli) from the vector/tensor multiplets due to having 8 supercharges.
Unlike in the previous Sections, the landscape of EFTS with 8 supercharges is much richer and the BPS mass gap highly depends on the  topological data characterizing each CY and entering through the value of the volumes.  In the next Section we will calculate explicitly the BPS mass gap in the case of 6d ${\cal N}=$1 theories by plugging this topological data from a list of known CY's. However, we can first provide some preliminary estimation of $\Lambda_{BPS}$ by assuming the simplest possible distribution of this data, in particular a diagonal topological intersection matrix as will be discussed next.
Intriguingly, this naive estimation of $\Lambda_{BPS}$ reproduces the results predicted by the species bound discussed in Section \ref{sec:species-2}, although it differs from the actual results computed numerically using explicit CY data in Section \ref{sec:8}. We show this coincidence with the species bound in the following, which seems to be rooted in the independence of the BPS states with the overall volume, and leave the comparison with the numerical results for Section \ref{sec:8}. 

We can organize the discussion according to the space-time dimension of the EFT as follows:

\begin{itemize}
\item 6d ${\cal N}=$1 theories (F-theory on elliptic CY$_3$)

The relevant BPS objects are strings arising from wrapping D3-branes on 2-cycles of the base of the elliptically fibered CY. Denoting $J$ as the Kahler form of the base, the tension of the strings read
\beq
T=\int_{\Sigma_2} J=q^I t_I
\eeq
where $q_I$ are the quantized charges and $t_I$ are the Kahler moduli defined as periods of the Kahler form over a suitable basis of 2-cycles in $H_2(B,\mathbb{Z})$. More details can be found in Section \ref{sec:8}. In order to get the tension in Planck units, we need to divide over the volume of the base or, equivalently, set the base volume to one. Hence, the above formula provides the string tension in Planck units subject to the constraint
\beq
\label{cons1}
\mathcal{V}=\frac12 t_IC^{IJ}t_J=1
\eeq
where $C^{IJ}$ is the inverse of the intersection form matrix in $H_2(B,\mathbb{Z})$. 
If one assumes that the intersection matrix is diagonal (so the trace grows linearly in $N$, the dimension of this space, i.e. $h^{1,1}(B)=N$), the constraint \eqref{cons1} implies 
\beq
Nt^2\sim 1\rightarrow t\sim N^{-1/2}
\eeq
so that the BPS tension scales as $T\sim N^{-1/2}$. Hence, the energy scale associated with this naive estimation of the tension is
\beq
\Lambda_{BPS}\simeq T^{1/2}\sim N^{-1/4}
\eeq
which coincides with the species bound in 6d, as given in \eq{species}.

\item 5d ${\cal N}=$1 theories (M-theory on CY$_3$)

The field content is organized into a gravity multiplet, vector multiplets and hypermultiplets. In total there are $h^{1,1}$ vectors but since one combination of them controlling the volume belongs to the hypermultiplets, there are only $h^{1,1}-1$ vector multiplets, whose scalar components parametrize the BPS masses. The vector multiplet moduli space can then be parametrised by $h^{1,1}$ real fields subject to the constraint that the overall volume is constant. In this case, the lightest BPS states are particles arising from wrapping M2-branes on real 2-cycles and BPS strings which come from M5 branes wrapping 4-cycles. Let us first focus on the M2 branes.  Their mass in Planck units is given by
\beq
\label{BPS2}
m_{BPS}= \int_{\Sigma_2}J=q_It^I
\eeq
where in addition we need to set the overall volume to one,
\beq
\label{cons2}
\mathcal{V}=\frac16 C_{IJK}t^I t^Jt^K=1
\eeq
Here,  the $t^I$ are the Kahler moduli, parameterizing the Kahler form as $J= t^I\omega_I$,
$\omega_I \in H^{1,1}(CY)$, the $C_{IJK}$ are the triple intersection numbers of the divisors $D_I$ dual to the $\omega_I$ and the number of moduli is $N=h^{1,1}(CY_3)$. Assuming again that the intersection matrix is diagonal, the constraint \eqref{cons2} implies
\beq
Nt^3\sim 1\rightarrow t\sim N^{-1/3}
\eeq
By plugging this into \eqref{BPS2}, the BPS mass scales as
\beq
\Lambda_{BPS}\sim N^{-1/3}
\eeq
which coincides again with the species bound in 5d.

As far as M5 branes wrapping 4-cycles, the tension of the resulting string is given by
\begin{equation}T_{BPS}=\int_{\Sigma_4} J\wedge J=\frac12 C_{IJK}t^I t^J q^K\end{equation}
where $q^K$ captures the charge class of the 4-cycle.  We thus have
\begin{equation}T_{BPS}\sim t^2\sim N^{-2/3}\rightarrow \Lambda_{BPS}\sim N^{-1/3}\end{equation}
We thus learn that both the M2 branes and M5 branes lead to the same naive $N$ dependence for the cutoff for large $N$ and both agree with the species bound in 5d.

\item 4d ${\cal N}=$2 theories (Type II on CY3)

The field content includes the gravity multiplet, vector multiplets and hypermultiplets. Again, the BPS central charges are determined by the scalars on the vector multiplets, which correspond to Kahler moduli of Type IIA on a $CY_3$ or to the complex structure moduli of Type IIB on the mirror $CY_3$. A similar classical analysis to the above for Kahler moduli sector would naively give a result that does not match with the species bound. However, the naive Kahler moduli space of Type IIA on CY$_3$ receives quantum corrections and it is better to compute the BPS masses in the complex structure sector of the mirror Type IIB theory which receives no corrections and where the following exact BPS mass formula holds for D3 branes wrapped around 3-cycles:
\beq
\label{BPSD3}
m_{BPS}= \frac{1}{{[\int \Omega \wedge \bar \Omega}]^{\frac12}}\int_{\Sigma_I}\Omega=\frac{q^I\Pi_I}{[{\Pi_I\bar \Pi_{\bar J}- \bar \Pi_I \Pi_{\bar J}}]^{\frac12}}
\eeq
 Here, $\Pi_I$ are the periods of the $(3,0)$-form of the CY over a symplectic basis of 3-cycles, and are holomorphic functions on the complex structure moduli. Note that $m_{BPS}$ is scale invariant and we can set the scale of $\Omega$ to one without loss of generality:
\beq
\int \Omega\wedge \bar \Omega=\Pi_I\bar \Pi_{\bar J}- \bar \Pi_I\Pi_{\bar J}=1
\eeq
Assuming that there are no significant cancellations and the number of terms is again given by  $N=h^{2,1}$, this suggest that $N\,\Pi^2\sim 1$, implying that the BPS mass scales as
\beq
m_{BPS}\sim N^{-1/2}
\eeq
which coincides with the species bound in 4d.

\end{itemize}

To sum up, by assuming a diagonal intersection matrix, where all nonzero entries have the same value, we get an estimation for the maximum BPS mass gap which coincides with the species scale in each dimension. This is very intriguing as the black hole arguments behind the species bound do not have direct information about the BPS spectrum of the theory. We will check this with explicit CY data in the next Section, finding that the actual growth of $\Lambda_{BPS}$ with $1/N$ is even bigger.

\section{BPS mass gap in 6d \texorpdfstring{${\cal N}=$}{N=}1 EFTs}\label{sec:8}
Having explored the desert scenario in compactifications with sixteen supercharges and after giving a heuristic derivation of species bound for the case of 8 supercharges, we now move to make more precise computations for the latter case.

 Although the philosophy and many of the details are very similar to the case of sixteen supercharges, with eight supercharges we encounter, for the first time, a large landscape of solutions, related to the existence of a large number of Calabi-Yau manifolds. By contrast, minimally supersymmetric compactifications in $d>6$ are all related to M theory on $K3$ \cite{Bedroya:2021fbu}, possibly with frozen singularities, which makes the number of possibilities very limited.

As already noted in the previous Section, compactifications with 8 supercharges include 4d $\mathcal{N}=2$, 5d $\mathcal{N}=1$ and 6d $\mathcal{N}=1$ models. Their stringy descriptions are type II on a CY$_3$,  M-theory on a CY$_3$, or F-theory on an elliptically fibered CY$_3$, respectively. In this Section we will address the question of gap only for 6d $\mathcal{N}=1$ models.  The reason for this restriction is a technical limitation, which we describe in more detail below. 

\subsection{Review of 6d \texorpdfstring{${\cal N}=$}{N=}1 and CY\texorpdfstring{$_3$}{3}'s with a toric base}
A 6d $\mathcal{N}=1$ supergravity includes a number of tensor, vector, and hypermultiplets \cite{polchinski1998string}. The F-theory model is an elliptic fibration over a complex surface base $B$, which is the actual compactification manifold of IIB string theory, and the degenerations of the fiber describe the location and types of 7-branes. Dimensional reduction of the self-dual 4-form $C_4$ of IIB on the 2-cycles of the base $B$ gives rise to the 6d tensor fields \cite{Vafa:1996xn}; because the signature of the intersection paring of $B$ is $(1,T)$ for some number $T$ (this is fixed by demanding that we have a $\mathcal{N}=1$ model \cite{Kumar:2010ru}), this reduction produces one anti-self dual tensor, which is part of the gravity multiplet, and $T=h^{11}(B)-1$ tensor multiplets. Vectors arise from the worldvolume degrees of freedom of 7-branes, and neutral hypermultiplets arise from the 7-brane moduli as well as the volume of the base, totaling $h^{21}(\text{CY}_3)+1$ hypermultiplets \cite{Vafa:1996xn}. Additional charged hypers can arise at self-intersections of 7-branes.

We want to explore the desert scenario in this class of models -- that is, we want to figure out how large can the gap be for the first heavy state. Just as before, there are many non-BPS states whose masses we cannot compute reliably: towers of KK modes, massive string excitations between branes, even the mass of charged states under vectors are unreliable because they cannot be BPS. There is a central charge in the 6d $\mathcal{N}=1$ algebra corresponding to a string charge, representing the tension of the BPS strings charged under the tensor multiplets. Given the tension of a BPS string $T$, we can estimate the mass of a closed string loop as
\begin{equation} m\sim \sqrt{T}.\label{mingap}\end{equation}
This will be our estimate for the mass gap of the theory at a generic point in moduli space. The actual mass gap may be lower, but certainly not higher than this estimate. Thus, the BPS strings are an upper bound for the actual mass gap. We should note that, just like in the 16 supercharges case, some of the BPS strings that we consider actually become tensionless at finite-distance singularities in the tensor moduli space, corresponding to SCFT points \cite{Katz:2020ewz}. By contrast, there are other strings, called supergravity strings in \cite{Katz:2020ewz}, where this does not happen.  Including or excluding these SCFT strings from our considerations will lead to two different notions of maximum gap: when strings are included, the gap describes the regime of validity for the low-energy free EFT that describes the dynamics at a generic point in moduli space; when they are excluded, the gap is set by purely gravitational objects as we keep the SCFT strings modes as part of the EFT. 

From the point of view of F-theory, the strings arise from wrapping $D3$ branes on 2-cycles (divisors) of the base $B$. The lattice of quantized string lattices $H_2(B,\mathbb{Z})$ is self-dual (but not necessarily even \cite{Seiberg:2011dr}).  $B$ is a complex manifold with vanishing odd Betti numbers, and admits a Kahler metric, and the volumes of calibrated 2-cycles are controlled by the Kahler form of the base $J_B\in H^2(B,\mathbb{R})$. Picking a two-cycle $\mathcal{C}_2\in H_2(B,\mathbb{Z})$ the tension of the corresponding string is
\begin{equation} T(\mathcal{C}_2)=\int_{\mathcal{C}_2} J_B.\label{BPSmass0}\end{equation}
Picking an arbitrary basis of $H_2(B,\mathbb{Z})$, which we denote $\{\alpha_I\}$, the Kahler form may be parametrized by the real Kahler parameters
\begin{equation} t_I\equiv \int_{\alpha_I} J_B. \label{kahlerdef}\end{equation}
Since the generic charge vector in the string lattice is given by an integer linear combination of the generators, $q^ I \alpha_I$ with $q^I\in\mathbb{Z}$, the BPS mass formula \eq{BPSmass0} can be rewritten as
\begin{equation} T=q^I t_I.\label{BPSmass1}\end{equation}
Not every string charge admits a BPS representative, but any two BPS strings (which preserve the same supercharges) are mutually BPS, and so the set of charge sites admitting a BPS representative naturally forms a cone. Thus, there is a cone over the integers, generated by some charge vectors $\{q^{\text{BPS},I}\}$, and since the tension of any BPS state must be nonnegative, we have
\begin{equation} q^{\text{BPS},I} t_I>0,\quad \text{for all }q^{\text{BPS},I}.\label{positiveT}\end{equation}
Since BPS states correspond to linear combinations of irreducible complex curves in the base, the cone of BPS states is actually the \emph{Mori cone} of the complex surface $B$. The conditions \eq{positiveT} define a dual cone in $J_B\in H^2(B,\mathbb{R})$, the Kahler cone. At boundaries of the Kahler cone, it can happen that a string becomes tensionless, signifying the appearance of an interacting SCFT. 

The Kahler parameters $t_I$ are dimensionful, but to study the mass gap in BPS particles, one typically works in six-dimensional Planck units. The 6d and 10d Planck masses are related by the volume of the surface $B$, which sits in one of the hypermultiplets in the F-theory description \cite{Vafa:1996xn}.  So to compute the tension in Planck units we can just divide \eq{BPSmass1} by the square root of the volume of the base,
\begin{equation} \mathcal{V}=\frac{1}{2}\int_B J\wedge J=\frac{1}{2}C^{IJ}t_I t_J,\label{vol}\end{equation}
where $C^{IJ}$ is the inverse of the intersection form matrix $C_{IJ}$ in $H_2(B,\mathbb{Z})$. Due to the large amount of supersymmetry, hypermultiplet and tensor multiplets do not mix, and can be tuned independently. Thus, equivalently, we can just work with the constraint that the internal volume is set to 1,
\begin{equation}\frac{1}{2}C^{IJ}t_I t_J=1.\label{internalvol}\end{equation}
While working with a Planck-sized compactification space does not make much sense physically, and indeed unprotected quantities are completely unreliable, the BPS spectrum is insensitive to these details.

To address the question of the BPS maximal gap, we need to know the Mori cone generated by the $\{q_a^{\text{BPS},I}\}$. This is a difficult question in general, and we do not know the answer for generic threefolds, or even generic elliptically fibered ones.  This is why we chose not to look at general 5d $\mathcal{N}=1$ or 4d $\mathcal{N}=2$ models. However, for those elliptic threefolds where the base $B$ is toric, the full answer was provided in \cite{Morrison:2012js}. This reference enumerated all possible toric bases for 6d F-theory models, and there is roughly 60000 of them; so toric bases provide a relatively large ``sub-landscape'' of 6d $\mathcal{N}=1$ compactifications where we can answer the question. 

As explained in \cite{Morrison:2012js}, the intersection structure and Mori cone of a toric surface with a given nonvanishing Hodge number $h^{11}(B)$ is fully described in terms of an overcomplete set of $h^{11}(B)+2$ divisors $\beta_K$, with intersection form $C_{KL}=\beta_K\cdot\beta_L$, where
\begin{equation} \beta_K\cdot \beta_{K+1}=\beta_K\cdot\beta_{K-1}=1,\quad \beta_K\cdot\beta_L=0\quad\text{otherwise}.\label{eww2}\end{equation}
The entries $\beta_K\cdot\beta_K$ are unconstrained and can have either sign, although at large $h_{11}$ we will find that the entries are mostly negative. The Mori cone is generated by the dual cone spanned by the $\{\beta_K\}$. We can now choose $h^{11}(B)$ elements of this list, say the first $h^{11}(B)$ entries, to construct the dual basis $\{\alpha_I\}$ above, and the associated Kahler parameters $t_I$. Using duality, we can also write the Kahler form as
\begin{equation} J(q)= t^I\alpha_I\cdot(q^K\beta_K)= q^KC_{KI}t^I,\end{equation}
where we have defined the dual Kahler parameters as
\begin{equation} t^I=C^{IJ}t_J.\end{equation}
In terms of the $t^I$, the volume of the $h^{11}(B)+2$ curves that generate the Mori cone is simply
\begin{equation}\mathcal{V}_K=C_{KI}t^I,\label{vols}\end{equation}
and the condition \eq{positiveT} is simply that $\mathcal{V}_K\geq 0$.

\subsection{Minimization algorithm and results}
We are now ready to explain the algorithm we will follow to find the maximal BPS gap. The starting point is equation \eq{vols} for the volumes. We wish to maximize the minimum of the $\mathcal{V}_K$ over all generators of the Mori cone, as a function of the $t^I$, subject to the constraint that the total volume is 1, \eq{internalvol}. Because the curve volumes depend homogeneously on the Kahler parameters $t^I$, we can always rescale them such that e.g. the smallest volume curve has any volume that we wish, say 1. The total volume of the base $\mathcal{V}$ is no longer one and the volume of the minimal curve before rescaling is related to the base volume $\mathcal{V}$ after rescaling as
\begin{equation} \text{min}(\mathcal{V}_K)\vert_{\mathcal{V}=1}=\frac{1}{\sqrt{\mathcal{V}\vert_{\text{min}(\mathcal{V}_K=1)}}}.\label{eew}\end{equation}
Our original problem, which is maximizing the left hand side of \eq{eew}, can then be traded by the dual problem of minimizing the right hand side. This is constrained optimization of a single function, as opposed to minmax optimization of $h^{11}(B)+2$ functions in the original problem.

In turn, this simplified problem can be solved exactly. The condition that the volume of all curves is $\geq 1$ defines a subcone of the Kahler cone, called the \emph{stretched Kahler cone} \cite{Demirtas:2018akl}, and so our problem is minimizing the quadratic function 
\begin{equation}\mathcal{V}=\frac12 C_{IJ} t^I t^J\end{equation}
 over a convex cone. The minima are located either at where $\nabla\mathcal{V}$ vanishes, on the boundary of the cone or at infinity. Since
 \begin{equation} \nabla\mathcal{V}\cdot q= C_{IJ} t^I q^J\end{equation}
 is the volume of a curve, and all the curve volumes are $\geq1$, the gradient never vanishes. Furthermore, since the volume of the curves are positive (i.e. they are effective), $\nabla \mathcal{V}>0$, implying that the base volume decreases as the volume of the curves decreases. Hence, the minimum of $ \mathcal{V}$ is located on the boundary of the stretched Kahler cone, where the curves take their minimum allowed value. Since the cone is defined by the condition
 \begin{equation} C_{KI}t^I\geq1\quad \forall\, K\end{equation}
 the boundary is piecewise linear, and each face is defined by setting the volume of one of the  $h^{11}(B)+2$ defining curves to 1. Within this face, we can run the argument again, to conclude that the minimum is located at another point where a second curve is set to 1, and proceeding iteratively, we conclude that the minimum is found at a point where the volumes of $h^{11}(B)$ curves are set exactly to one.
 
 Once the volumes of $h^{11}(B)$ curves are set, the Kahler parameters $t^I$ are fully determined, and so are the volumes of the remaining 2 curves. This leads to the following simple algorithm to find the minimum:\begin{enumerate}
 \item For each of the 
 \begin{equation} \binom{h^{11}(B)+2}{h^{11}(B)}=\frac{(h^{11}(B)+2)(h^{11}(B)+1)}{2}\end{equation}
 possible choices of $h^{11}(B)$ curves, find the value of the parameters $t^I$ that set all of their volumes to 1.
 \item Find the volumes of the additional 2 curves. If either of these volumes is below 1, the corresponding $t^I$ lies outside of the stretched Kahler cone, and we ignore it. Otherwise, the point is kept, and we compute the corresponding $\mathcal{V}$.
 \item The global minimal volume corresponds to the minima of all the values of $\mathcal{V}$ found in the previous point.
 \end{enumerate}
 
 We applied this algorithm to the list of 61539 toric bases constructed in \cite{Morrison:2012js}. We excluded examples including curves of self-intersection -9,-10 and -11, since as explained in \cite{Morrison:2012js} the corresponding bases are not toric after blow-up, and the cycle structure described above does not apply to these examples. This still leaves us with 34868 different explicit bases, which constitute the ``mini-Landscape'' where we performed our numerical experiments. 
 
 \begin{figure}[!htb]
\centering
\includegraphics[width = 0.7 \textwidth]{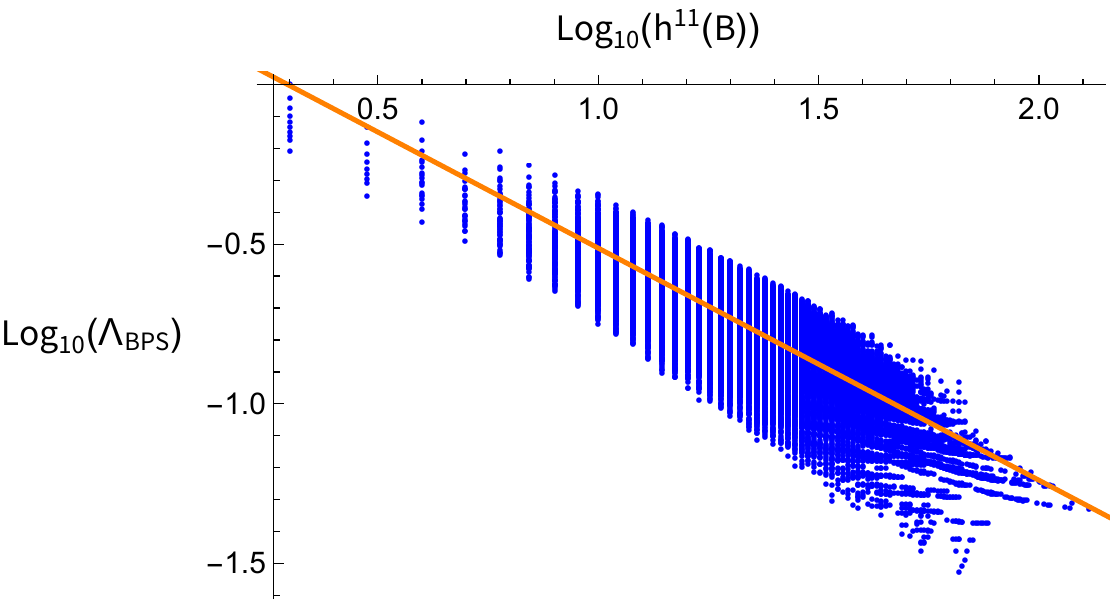}
\caption{Logarithmic plot of the minimal gap obtained from equation \eq{mingap} versus $h^{11}(B)$, for the 34868 toric bases in the toric mini-Landscape  described in \cite{Morrison:2012js}. There is some spread corresponding to different models with the same $h^{11}(B)$, but there is an overall linear trend, with the best fit shown, corresponding to the function $h^{11}(B)^{-n}$ with $n\sim 0.7$.}
\label{f1}
\end{figure}
 
The results are shown in Figure 2, where we plot $h^{11}(B)$ versus the minimal gap in Planck units (equivalently, the cutoff $\Lambda_{\text{BPS}}$ set by the BPS states), given by \eq{mingap} applied to the solution of the minmax problem in the left hand side of \eq{eew}. We see some spread in the data, and a general downwards trend, which is roughly linear in the logarithmic plot we provide; the best fit is the curve
\begin{equation}m_{\text{gap}}\sim\Lambda_{\text{BPS}}\sim\frac{1}{h^{11}(B)^n}.\label{w22}\end{equation}
with $n\sim 0.7$. The total volume of the base scales with $h^{11}(B)$ as $h^{11}(B)^{2n}=h^{11}(B)^{1.4}$, which will be relevant later.

 Unfortunately we cannot provide a sharp upper bound on $h^{11}(B)$, since this would also depend on the $\mathcal{O}(1)$ coefficients responsible for the twiddles in \eq{w22} and \eq{w33} which we do not control. We should also point out that it has been shown that the Landscape of elliptically fibered Calabi-Yau threefolds is finite, see \cite{gross1994finiteness}. This is consistent with our results, since the mini-Landscape we look at is contained within this class.
 
 At this point we would like to recall the general argument provided in Section 3, where with some genericity assumption, we concluded that the cutoff coming from the strings should saturate the species bound. The reason for the mismatch is that the genericity assumption we made, that $C_{IJ}$  is close to the identity matrix, is a bad one in our mini-Landscape. As shown in eq. \eq{eww2}, the actual intersection matrix we obtain has off-diagonal entries and diagonal values which can be far from 1. Writing the volume as
 \begin{equation}\mathcal{V}=\frac12 C^{IJ} t_I t_J,\end{equation}
 the vector $t_I$ of volumes has components all close to 1 at the minimum (and in fact, at least all components except for two are exactly one). The partial inverse matrix $C^{IJ}$ has many nonzero entries of all sizes, and is in fact far from the identity for a typical base. As a result, $C_{IJ} t^I t^J$ grows faster than $Nt^2$, resulting in a faster decrease of $t$ with $N$.

The failure of this argument should be viewed as a cautionary tale: It is dangerous to make general arguments about string compactifications, involving topological data such as $C_{IJ}$, or triple intersections $C_{IJK}$ using ``genericity'' assumptions without actually having concrete examples to back them up. 

 \subsection{Field theory vs. black hole mass gap and species bound}
 It is also very important to understand the nature of the BPS strings providing the gap above. Roughly speaking, the strings come in three different classes \cite{Katz:2020ewz}, according to the sign of the self-intersection $q\cdot q= q^I C_{IJ} q^J$:
 \begin{itemize}
\item Strings with $q\cdot q<0$ can become tensionless at finite distance singularities in moduli space, where they give rise to SCFT's. These are the analog of the field theory states described in Subsection \ref{sec:16}; in a sense, they are captured by (conformal) field theory description, and at a generic point in moduli space they are heavy only due to a field-theoretic Higgsing mechanism. They are also strongly interacting and self-repulsive: they do not form bound states at any point in moduli space, as the corresponding large charge states would be described by a CFT of negative central charge.
\item Strings with $q\cdot q>0$ are intrinsically gravitational objects, called \emph{supergravity strings} in \cite{Katz:2020ewz}. Unlike their field-theory counterparts, they can form bound states, and for large charge are described by black strings of nonzero horizon area, with near-horizon geometry $AdS_3\times S^3$. They never become tensionless, and in fact, their tension is lower-bounded by 1 in Planck units.
\item Finally, there are states with $q\cdot q=0$, such as the fundamental heterotic string, which sit at the divide between field theory and gravity strings. For large values of the charge, these may be described by singular dilatonic zero-area solutions, such as the small black holes in \cite{Hamada:2021yxy}. Hence we may call them \emph{small supergravity strings}. Unlike honest supergravity strings, whose tension remains superplanckian throughout moduli space, small supergravity strings become tensionless at infinite distance limits in moduli space. The equivalent version in 4d theories were called \emph{EFT strings} \cite{Lanza:2021udy} and play an essential role when testing the Distance Conjecture \cite{Ooguri:2006in}.
 \end{itemize}
For $h_{11}>2$, the minimal volume curves have negative self-intersection, and so the limits presented in Figure \ref{f1} for $h_{11}$ above this value all correspond to deformations of SCFT strings where the strings have acquired maximal tension. The physical interpretation of the cutoff is then the scale at which the IR free EFT at generic points in moduli space breaks down (similar to the GUT scale in the GUT scenario), and is replaced by an interacting, but still non-gravitational description. Above this cutoff, the effective description involves several interacting light strings that are not mutually local (each of them becomes tensionless at different points in moduli space), and so we are unlikely to be able to describe them via a local field theory. 

\begin{figure}[!htb]
\centering
\includegraphics[width = 0.7 \textwidth]{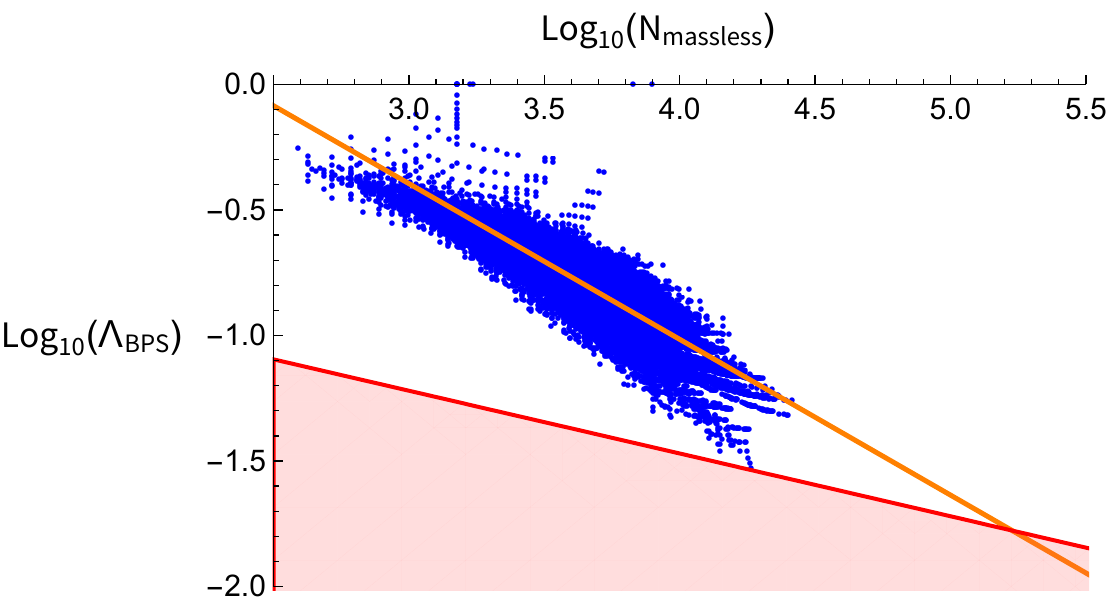}
\caption{Schematic plot where the toric examples are plotted on top of the restriction required by a qualitative version of the species bound, involving only massless fields. This naive expression would exclude situations where the lightest field is below the blue line, and so is not compatible with the linear trend observed in the data if this is extrapolated indefinitely, suggesting a maximum value for $N$. There is an uncontrolled $\mathcal{O}(1)$ factor in the species bound line, which manifests in this logarithmic plot as an uncontrolled offset. This is why the plot is only schematic, and we cannot make a sharp prediction for the maximal $N$. }
\label{f2}
\end{figure}

We can now compare this BPS gap with the prediction of the species bound. To do this, we compute the total number of massless modes (tensors, vectors and hypermultiplets) for each CY, by identifying the existing non-higgssable clusters and using anomaly cancellation conditions, and using the results of \cite{Morrison:2012js} to count the number of vector and hyper multiplets. Each of these multiplets has eight bosonic degrees of freedom. Taking this into account, the result is plotted in Figure \ref{f2} where the best fit reads
\beq
\Lambda_{\rm BPS}\sim \frac{1}{N^{0.62}}\label{w332}
\eeq
Intriguingly, the gap associated to the string states decreases with $N$  faster than what the six-dimensional species bound would suggest,
\begin{equation}\Lambda_{\text{sp}}\sim\frac{1}{N^{\frac14}}.\label{w33}\end{equation}
As we argued in Section \ref{sec:species-2},  we can have points which are desert-like in which case the species bound cutoff should depend mainly on the number of exactly massless fields.  Assuming this we will now check if this leads to any contradictions between \eqref{w332} and \eqref{w33}.

We will also make the reasonable assumption that below the species scale $\Lambda_{\rm sp}$ there exists a consistent EFT coupled to gravity including all the states with mass less than this scale.  On the other hand we expect that in the string examples we studied the points which give the maximal gap cannot be described in the context of a consistent EFT as in that case we have no bounds in how large we can make the BPS tensions. By contrast, in an actual field theory description, with gravity decoupled, the vevs of tensor multiplets (and hence, the string tensions) can grow without bound.  Therefore this implies that $\Lambda_{BPS}\geq \Lambda_{\rm sp}$.

 Since the slope of $\Lambda_{\rm BPS}$ and $\Lambda_{\rm sp}$ are different, they are going to cross at some point. Whenever $\Lambda_{\rm BPS}$ is above the species bound, there is no conflict, as the mismatch can be accounted for in principle by lighter non-BPS states we cannot control, such as Kaluza-Klein modes, etc. Having the species bound above $\Lambda_{\rm BPS}$ would not only go against the desert scenario, but would also imply that we can have a field theory description above $\Lambda_{\rm BPS}$ 
 leading to a contradiction with $\Lambda_{BPS}\geq \Lambda_{\rm sp}$.
 Hence, the dependence of $\Lambda_{\rm BPS}$ we found experimentally from the mini-Landscape of toric bases is incompatible with this version of the species bound in terms of massless fields, unless the family is cut short at a maximal value of $N$, as illustrated in Figure \ref{f2}, so the region after the two curves cross is ruled out. In other words, absence of conflict with this strengthened version of the species bound provides yet another hint that the Calabi-Yau landscape is finite, by suggesting an upper bound of the number of massless fields and, consequently, on the  Hodge numbers.
 
It is also interesting to explore the gap set by purely gravitational objects.  To compute this gap, we must look at (small) supergravity strings. This can be addressed with an algorithm similar to the one outlined in the previous Section, but we can in fact achieve a shortcut based on the structure of the toric mini-Landscape we are studying. As explained in \cite{Morrison:2012js}, all toric bases we consider are obtained from blow-ups of either $\mathbb{CP}^2$ or one of the Hirzebruch surfaces $\mathbb{F}_m$, where $0\leq m\leq 11$, which hence constitute a set of ``minimal models''. For any given toric base $B$ we can always move to a point in the tensor moduli space which is arbitrarily close to the blow-down to one of the minimal models. When doing this no supergravity string becomes tensionless. This means that the global maximum of the minimal supergravity string tension will then be at least as large as the value obtained from one of the 13 minimal models described above\footnote{There is yet another way to define the quantum gravity cutoff namely that defined by the horizon size of the black strings, which we also explored. Since supergravity strings have a nonzero horizon area, we can estimate the cutoff of the EFT from the size of the smallest possible value of the area, as $\Lambda\sim (\text{Horizon area})^{-\frac{1}{3}}$. In turn, the horizon area is related to the central charges of the string \cite{Kim:2019ths}, which themselves depend on the intersection numbers and the products of the charge vector with the canonical class of the base. We explored this question via a Monte-Carlo algorithm and found that, for moderate $h^{11}(B)$, there is always a curve with horizon area of Planck size, with no noticeable dependence in $h^{11}(B)$. Thus, it seems that the cutoff estimated by this method is also $\mathcal{O}(1)$ in Planck units. }. Having reduced the analysis to these 13 cases, one can compute explicitly that the maximum gap is ${\mathcal O}(1)$, independently of $h^{11}(B)$, similar to what we had seen in the context of theories with higher supersymmetry.

\subsection{Relation to the Tadpole Conjecture}
Our results can also be of interest for other questions currently under scrutiny. For instance, there is a strong connection to the Tadpole Conjecture \cite{Bena:2020xrh,Bena:2021wyr}. This conjecture claims that, in vacua with a large number of moduli, the number of flux quanta that would be required to stabilize all of them is too large to satisfy the tadpole conditions. This yields an obstruction to stabilize all moduli and puts into question the existence of a large landscape of flux vacua. In the context of Type IIB Calabi-Yau flux compactifications, the conjecture claims that the flux contribution to the D3-brane tadpole condition scales as
\beq
N_{\rm flux}=\int F_3\wedge H_3> 2\alpha (h_{2,1}+1)\quad \text{for large }h_{2,1}
\eeq
where $h_{2,1}$ is the number of complex structure moduli and $\alpha$ is conjectured to be $1/3$. Reference \cite{Plauschinn:2021hkp} checked this conjecture in the large complex structure regime by using Kahler moduli data of the mirror dual Calabi-Yau threefold in the Kreuzer-Skarke database \cite{Kreuzer:2000xy} presented in \cite{Demirtas:2018akl}. The large complex structure regime maps to the stretched Kahler cone of the mirror defined by setting all curve volumes $\geq 1$ to keep computational control.  In this regime, minimization conditions of the self-dual flux potential imply that $N_{\rm flux}$ scales as the volume of the mirror plus some additional terms. Hence, checking the conjecture reduces to finding the minimal value of the volume within the stretched Kahler cone. This is exactly the same optimization problem that we have attacked in this paper! 
In  \cite{Plauschinn:2021hkp} this minimal volume was estimated, although not computed explicitly, using the results of \cite{Demirtas:2018akl} to grow as $N_{\rm flux}\gtrsim \text{min}(\text{vol}(CY_3))\sim h_{2,1}^{6.6}$. We cannot apply directly our results to this setup, since we have only computed explicitly the volume of the base of the elliptically fibered threefolds, and not of the total Calabi-Yau. However, assuming they scale similarly, we obtain a power much smaller than the one estimated in \cite{Plauschinn:2021hkp} (namely $\sim h_{2,1}^{1.4}$) which is consistent but slightly stronger than the Tadpole Conjecture. 

In the case of $d=5$ $\mathcal{N}=1$ EFTs the optimization problem to obtain the maximal BPS gap resembles even more the one required to check the Tadpole Conjecture at large complex structure in \cite{Plauschinn:2021hkp}, since it is the overall volume of the CY that enters in the BPS mass formula. Interestingly, the naive estimation of the BPS mass gap performed in Section \ref{sec:heuristic} using genericity arguments (yielding $\mathcal{V}_{\rm min}\sim N$) would precisely reproduce the scaling of the Tadpole Conjecture. However, the lesson from studying actual examples in 6d is that these genericity arguments can be too simplistic and  $\Lambda_{BPS}$ can grow even faster with the number of massless fields. We will promote this observation to the ``BPS desert conjecture'' in the Conclusions, which would imply that the Tadpole Conjecture is satisfied with room to spare. It should be emphasized that arguments of this kind refer to generic features of the Calabi-Yau landscape, and do not exclude the possibility of special cases that evade them.

We suspect that the connection with the Tadpole Conjecture runs deeper than what this analysis in the large complex structure regime (or the stretched Kahler cone of the mirror) suggests.  Recall that the maximum BPS mass gap occurs at a special point deep in the interior of the moduli space far away from all boundaries. Taking into account that $N_{\rm flux}$ for self-dual fluxes increases when approaching the boundaries \cite{Plauschinn:2021hkp}, it is natural to expect that the minimal value of $N_{\rm flux}$ will happen near (or perhaps even at the same) point that maximizes the BPS mass gap.  This provides further motivation to study these special points in more detail.

\section{Conclusions and finiteness of the CY landscape}\label{conclus}
In this paper we have explored the desert scenario in the string landscape; namely searching for points on moduli of supergravity theories where there are no massive states until mass scales close to the Planck scale.  We have found that for a number of examples the points on moduli where we obtain a maximal gap are well in the interior of the moduli space, and in a sense almost by definition among the most interesting and difficult points to analyze in the moduli space of supergravity theories.  As such, regardless of the motivation of the present paper, {\it it would be interesting to have a more systematic study of such special points in the string landscape}.  In addition we have focused on the mass scales set by BPS data which allow us to compute the maximal cutoff set by BPS states $\Lambda_{BPS}$ exactly.  Here we have found two surprises:  The first surprise is that the naive supersymmetric arguments for theories with 8 supercharges lead to the $\Lambda_{BPS}$ scaling with the number of massless modes $N$ exactly as predicted by the species scale.  The second surprise is that this naive argument is not consistent with a large class of actual examples which suggest that $\Lambda_{BPS}$ becomes too small too quickly with increasing $N$.  To avoid contradiction with the species scale we propose that there are massive states lower than $\Lambda_{BPS}$ not associated with BPS objects leading to $\Lambda <\Lambda_{BPS}$ which changes the slope of this dependence.  However, this can only work in a finite range of $N$ due to the fact that the BPS cutoff sets an upper bound for the cutoff of the full theory. This therefore leads to an upper bound on the number of massless modes on Calabi-Yau manifolds.  Turning this around, if we could find a theoretical reason behind the higher slope for the dependence of $\Lambda_{BPS}$ on the number of massless modes, it would lead to a theoretical explanation of the upper bound on the number of massless modes in these theories.

Motivated by this example we make the more general conjecture, the {\it BPS desert conjecture}, that $\Lambda_{BPS}\sim N^{-a}$ where $N$ is the number of massless modes with $a>{\frac{1}{d-2}}$ in all cases and in all dimensions for supergravity theories.   This would then give a theoretical reason for the existence of an upper bound on the number of massless modes in a supergravity theory, which is a requirement for finiteness of the string landscape.  Given that $\Lambda_{BPS}$ is in principle computable with known techniques it would be interesting to check the general BPS desert conjecture.   This conjecture also has other important implications.  For example considering the case of $d=5$ and applying it to the case of M-theory on arbitrary (not necessarily elliptic) CY 3-folds we obtain that if we fix the volume of holomorphic 2-cycles to be larger than 1, then the volume of CY will scale as $h_{11}^{b}$ with $b>1$ (since the volume of 2-cycles would scale as $N^{-a}$ with $a>\frac13$ if we fixed the overall volume to be 1) and this leads to a stronger form of the tadpole conjecture at large complex structure\footnote{It is natural to speculate that the 4d version of the BPS desert conjecture will also lead to a similar bound for an arbitrary point in complex structure in flux compactifications.} using the setup of \cite{Plauschinn:2021hkp}.

It is exciting to try to apply the general lessons we have learned from this supersymmetric study to the Standard Model, even though we have to be cautious due to the lack of supersymmetry in the real world.  One of the typical scenarios considered plausible for the SM is precisely the desert scenario, where there are no extra massive states other than SM fields until we reach the GUT scale.  At the GUT scale we get new massive states which include massive fields needed to complete the Standard Model to a GUT (perhaps in a higher dimensional sense as is often the case in string theory \cite{Beasley:2008dc}).   Viewing the Standard Model states as the analog of massless states, in a sense this is very similar to what we have found at special interesting points in the moduli of the supersymmetric theories,  where the BPS gap or the species scale are set by massive field theory states.  We can push the massive states $\Lambda$ very close to the Planck scale, differing from it by $N^{-a}$ where $N$ is the number of massless modes and $a>\frac{1}{d-2}$ which for $d=4$ leads to $a>1/2$.  Taking into account that the number of massless modes for the Standard Model is about $N\sim 10^2$, this suggests that $\Lambda_{GUT}< 10^{-1} M_{\text{P}}$. In particular if we had $a\sim1$ we would rather be close to the expected value of $\Lambda_{GUT}\sim 10^{-2} M_{\text{P}}$. An alternative explanation of the mismatch could be the existence of a dark sector with a large number of degrees of freedom, with $N\sim 10^4$, that would also lower the scale.
 
\section*{Acknowledgments}
We would like to thank Severin Luest for useful discussions. We are grateful to the Simons Center for Theoretical Physics, Stony Brook, for hospitality during the early stages of this work in the context of its Summer Workshop. The work
of MM, CV is supported by a grant from the Simons Foundation (602883, CV) and by the
NSF grant PHY-2013858, which also supported IV in the early stages of this work. The work of IV is now supported by the grants RYC2019-028512-I from the MCI (Spain), PGC2018-095976-B-C21 from MCIU/AEI/FEDER, UE and the grant IFT Centro de Excelencia Severo Ochoa SEV-2016-0597.

\bibliographystyle{JHEP}
\bibliography{refs-finiteness-CY}

\end{document}